\documentclass[preprint,showpacs,preprintnumbers,amsmath,amssymb]{revtex4}
\usepackage{graphicx}
\usepackage{epsfig}
\usepackage{bm}
\usepackage{amsfonts}
\usepackage{subfigure}
\usepackage{multirow}
\usepackage{float}
\usepackage{color}
\providecommand{\tabularnewline}{\\}

\begin{document}

\title{\textbf{Primordial black holes in non-canonical scalar field inflation driven by quartic potential in the presence of bump}}

\author{Soma Heydari\footnote{s.heydari@uok.ac.ir} and Kayoomars Karami\footnote{kkarami@uok.ac.ir}}
\address{\small{Department of Physics, University of Kurdistan, Pasdaran Street, P.O. Box 66177-15175, Sanandaj, Iran}}
\date{\today}

%============================================Abstract===================================================
\begin{abstract}
\noindent
Here, generation of Primordial Black Holes (PBHs) from quartic potential in the presence of a tiny bump in non-canonical inflationary model has been inquired. It is demonstrated that, a viable inflationary era can be driven through the quartic potential in non-canonical framework with a power-law Lagrangian density. Furthermore, setting a suitable function of inflaton field as a correction term (like a bump) to the quartic potential, causes the inflaton to slow down for a while. In such a short time span, the amplitude of the scalar perturbations power spectrum on small scales grows up sufficiently versus CMB scales. In addition to the bump feature, the enhancing effect of the $\alpha$ parameter of the  Lagrangian on the amplitude of the scalar power spectrum has been shown. Fine tuning of three parameter Cases of the model results in generating of three Cases of PBHs. In addition, we investigate the secondary Gravitational Waves (GWs) produced during generation of PBHs and show that their contemporary density parameter spectra $(\Omega_{\rm GW_0})$  can be tracked down by GWs detectors.
 \end{abstract}
\maketitle

%============================================Introduction===================================================
\newpage
\section{Introduction}
It is well acknowledged that, primal curvature perturbations can be generated in inflationary epoch. By reason of  accelerated expansion of the cosmos during the inflationary era, the pertinent scales to these perturbed modes  exit the Hubble horizon. Reentry of the scales with enough sizable amplitude to the horizon gives rise to produce the ultra-condensed localities in  Radiation Dominated (RD) era. In due time, generation of Primordial Black Holes (PBHs) originates from  gravitational cave-in of these localities. The notion of PBHs was introduced in the 1970s  \cite{zeldovich:1967,Hawking:1971,Carr:1974,carr:1975} at the very earliest. Thereafter, advantageous discovery of Gravitational Waves (GWs) emanated from two coalescing  black holes with masses around  $ 30 M_\odot$ ($M_\odot$ connotes the solar mass) by LIGO-Virgo teamwork \cite{Abbott:2016-a,Abbott:2016-b,Abbott:2017-a,Abbott:2017-b,Abbott:2017-c}, has renovated curiosity about PBHs notion. furthermore, puzzling essence of Dark Matter (DM) content of the universe \cite{Garrett:2011} and inability to observe the particle DM, have encouraged the scientific community to ponder about PBHs as a remarkable source for the entirety or a fraction of DM content and detected GWs
\cite{Ashrafzadeh:2023,Wang:2023,Khlopov:2010,
OGLE-1,Kamenshchik:2019,fu:2019,Dalianis:2019,mahbub:2020,mishra:2020,Solbi-a:2021,Solbi-b:2021,Teimoori-b:2021,Laha:2019,Teimoori:2021,fu:2020,
Dalianis:2020,Pi,Garcia-Bellido:2017,Heydari:2022,Heydari-b:2022,Rezazadeh:2021,Kawaguchi:2023,Kawai:2021edk,Saridakis:2023,Clesse:2015,Kawasaki:2016,
Braglia:2020,Motohashi:2017,Sayantan-4:2023,Sayantan-6:2023,
Cai:2023,Belotsky:2014,Ahmed,Belotsky:2019,Domenech:2021,Domenech-1:2020,Domenech-2:2020,Drees:2021,Kawai:2022emp,Yuan:2021}.

Owing to the non-stellar origin of PBHs generation, they could be located in a wide permitted mass band.
PBHs with mass  ${\cal O}(10^{-5})M_\odot$  and abundance  ${\cal O}(10^{-2})$, can be located in the permitted span of OGLE data  \cite{OGLE-1,OGLE-2} and they can be suitable to expound the ultrashort-timescale microlensing events in this zone.
Due to the fact that, There is no constraint on PBHs abundance in the mass scope of ${\cal O}(10^{-16}-10^{-11})M_\odot$, the located PBHs in this zone could be contemplated as an appropriate candidate for the DM content in its totality \cite{Teimoori:2021,Teimoori-b:2021,Heydari:2022,Heydari-b:2022,Rezazadeh:2021,fu:2019,Dalianis:2020,Solbi-a:2021,Solbi-b:2021}.

It is demonstrated that, amplification of the amplitude of the curvature perturbations power spectrum to order ${\cal P}_{\cal R}\sim{\cal O}(10^{-2})$ on small scales is indispensable to generate discernible PBHs
\cite{Ashrafzadeh:2023,OGLE-1,Kamenshchik:2019,fu:2019,Dalianis:2019,mahbub:2020,mishra:2020,Solbi-a:2021,Solbi-b:2021,Teimoori-b:2021,Laha:2019,Teimoori:2021,
fu:2020,Dalianis:2020,Pi,Garcia-Bellido:2017,Motohashi:2017,Heydari:2022,Heydari-b:2022,Rezazadeh:2021,Kawaguchi:2023,Kawai:2021edk,Saridakis:2023}.
The imposed confinement as ${\cal P}^{*}_{\cal R}\simeq2.1 \times 10^{-9}$  at pivot scale $k_{*}=0.05~ \rm Mpc^{-1}$ by CMB observational constraint  \cite{akrami:2018}, could be a challenge to amplify the ${\cal P}_{\cal R}$ to such a mentioned order on small scales.
Hitherto, numerous methodologies have been proposed to enhance the curvature power spectrum about seven order of magnitude on small scales relative to CMB scales. This aim could be achievable in Standard Model (SM) of inflation \cite{Dalianis:2019,mahbub:2020,Garcia-Bellido:2017,Motohashi:2017,mishra:2020}, modified gravity models \cite{Ashrafzadeh:2023,Kawaguchi:2023,Kawai:2021edk,Pi}, scalar- tensor theories \cite{Dalianis:2020,fu:2019,Teimoori:2021,Heydari:2022,Heydari-b:2022,fu:2020} and non-canonical scalar field models \cite{Kamenshchik:2019,Saridakis:2023,Solbi-a:2021,Solbi-b:2021} as well.
Over and above the single field inflationary models, enhancement of the ${\cal P}_{\cal R}$ and production of PBHs in multi-field inflationary models could be achievable \cite{Clesse:2015,Kawasaki:2016,Braglia:2020,Kawai:2022emp}.
The additional effect of returning the amplified curvature perturbation to the horizon in RD stage, would be the propagation of secondary gravitational waves synchronous to PBHs generation
\cite{Kamenshchik:2019,fu:2020,Dalianis:2019,mahbub:2020,Solbi-a:2021,Solbi-b:2021,Teimoori-b:2021,Laha:2019,Teimoori:2021,fu:2020,Dalianis:2020,
Heydari:2022,Heydari-b:2022,Rezazadeh:2021,Kawaguchi:2023,Kawai:2021edk,Saridakis:2023}. Accompanied GWs with PBHs could be traceable if their density parameter spectra lie in the sensitivity scope of GWs detector such as Square Kilometer Array (SKA) \cite{ska}, European Pulsar Timing Array (PTA) \cite{EPTA-a,EPTA-b,EPTA-c,EPTA-d}, Laser Interferometer Space Antenna (LISA) \cite{lisa,lisa-a}  and so forth. Subsequently, the credibility of PBHs models could be inspected in light of the observational data of GWs detectors.

The notable point about PBHs generation scenarios is that, the inflation could occur during a multi-phase procedure in most of them.
Mostly in these models, it has been tried to slow down the inflaton field in a transitory span of friction governed domain, to wit Ultra Slow Roll (USR) inflation, vis a vis the Slow Roll (SR) domain to provide the enough time for ${\cal P}_{\cal R}$ to grow up sufficiently. In USR phase the SR condition is contravened and appropriate conditions are provided for enhancing the ${\cal P}_{\cal R}$. Subsequently, in this way one can certify that the detectable PBHs will be able to produce after inflation in RD era.
In the SM of inflation, the USR span to produce PBHs could be attained either by dint of an inflection point in the inflationary potential \cite{Dalianis:2019,mahbub:2020,Garcia-Bellido:2017,Motohashi:2017}, or adding a correction term like a bump/dip to the base potential to reduce the velocity of the inflaton \cite{mishra:2020,Rezazadeh:2021}.
In \cite{Dalianis:2020,fu:2019,Teimoori:2021,Heydari:2022,Heydari-b:2022,fu:2020}, the PBHs generation has been investigated in NonMinimal Derivative Coupling (NMDC) to gravity  framework as a subclass of Horndeski theory \cite{Defelice:2013}. In these models choosing of proper function of scalar field as a coupling parameter between field derivative and gravity gives rise to enhance the scalar power spectrum and produce detectable PBHs in RD era.

 Furthermore, PBHs generation in non-canonical scalar fields model with a power-law form of Lagrangian applying a steep-deformed exponential potential  has been inspected in  \cite{Saridakis:2023}. It is well known that, the non-canonical scalar field inflationary model with power-law Lagrangian ${\cal L}(X,\phi)=X^{\alpha}-V(\phi)$ could be a conceivable generalization of the SM of inflation \cite{Unnikrishnan:2012,Rezazadeh:2015,Mishra:2022} ($X$ is the canonical kinetic  term). In this model $\alpha$ parameter denotes deviation from  canonicity, in the other statement for  $(\alpha=1)$  canonical Lagrangian of the SM of inflation is revived. It has been proved that the SR parameters, using the mentioned form of Lagrangian, could decrease  in comparison with the canonical case. Thus the SR conditions could be achieved more comfortably and the length of inflation could be increased. So the larger scalar spectral index $(n_{s})$ and smaller tensor-to-scalar ratio $(r)$ rather than the ones in canonical case could be attained \cite{Unnikrishnan:2012}. As a result such a non-canonical scalar field model can be a useful framework to reanimate the ruled out steep potentials like exponential, power-law and inverse power-law  potentials \cite{Unnikrishnan:2012,Rezazadeh:2015,Mishra:2022} in light of recent observational data of Planck 2018 \cite{akrami:2018}.

A feasible way to accommodate  the quartic potential to the viable inflation, is applying the power-law non-canonical framework \cite{Unnikrishnan:2012,Mishra:2022}. So, in this work, we are interested in studying the generation of PBHs and GWs in the non-canonical setup containing power-law Lagrangian, in which the quartic potential with a correction term like a bump has been embedded. Furthermore, we try to show the amplifying effect of the $\alpha$ parameter of the non-canonical Lagrangian on the enhanced scalar power spectrum on small scales to produce PBHs.

This paper is arranged as follows. Firstly, basic groundwork of non-canonical scalar field model with power-law Lagrangian is reviewed in Sec. \ref{sec2}. Thence, the procedure of amplifying the amplitude of the curvature perturbations at small scale in our setup has been described in Sec. \ref{sec3}. From then on Sec. \ref{sec4} is devoted to compute the mass spectra of the  generated PBHs. At last the current density parameter spectra of secondary GWs are analyzed  in Sec. \ref{sec5} and the main consequences of our work are epitomized in Sec. \ref{sec6}.
%==========================================non-canonical model======================================================
\section{groundwork of non-canonical power-law setup}\label{sec2}
we start with the following generic action
\begin{equation}\label{action}
S=\int{\rm d}^{4}x\ \sqrt{-g}\ {\cal L}(X,\phi),
\end{equation}
wherein the Lagrangian density $ {\cal L}(X,\phi)$ can be designated as various functional forms of scalar field $\phi$ and kinetic term $X=\frac{1}{2}\partial_{\mu}\phi\partial^{\mu}\phi$ \cite{Panotopoulos-2007,Chimento-2004,Unnikrishnan:2012,Rezazadeh:2015,Mishra:2022}.
In this work, power-law form of the Lagrangian density is assumed
\begin{equation}\label{Lagrangian}
{\cal L}(X,\phi) = X\left(\frac{X}{M^{4}}\right)^{\alpha-1} - V(\phi),
\end{equation}
in which, $\alpha$ parameter is dimensionless, $M$ parameter with dimension of mass is related to scales having the non-canonical status  and $V(\phi)$ denotes the scalar field potential.
Such an inflationary model with non-canonical kinetic term is well studied in literature under the name of k-inflation too \cite{Barenboim}.
 As mentioned previously, $\alpha$ determines the diversion of canonicity, in other words for $\alpha=1$ the canonical Lagrangian ${\cal L}(X,\phi) = X - V(\phi)$ can be restored from Lagrangian (\ref{Lagrangian}).

In this section,  the main equations governing background and perturbations dynamics through inflationary non-canonical model driven by action (\ref{action}) with Lagrangian (\ref{Lagrangian}) have been reviewed. So the  Friedmann-Robertson-Walker (FRW) metric
for the homogeneous and isotropic cosmos is considered as
\begin{equation}
\label{eq:FRW}
{\rm d}{s^2} = {\rm d}{t^2} - {a^2}(t)\left( {{\rm d}{x^2} +
{\rm d}{y^2} + {\rm d}{z^2}} \right),
\end{equation}
wherein  $a(t)$ and $t$  designate the scale factor and cosmic time. Applying the FRW metric, the kinetic term takes the form $X ={\dot \phi ^2}/2$ (the dot is derivative versus $t$).
For the Lagrangian (\ref{Lagrangian}),  the energy density $(\rho _\phi)$ and pressure $(p_\phi)$ of the scalar field  can be obtained as (see \cite{Unnikrishnan:2012} to review these equations in details)
\begin{eqnarray}
{\rho _\phi } &=& \left( {2\alpha - 1}
\right)X{\left( {\frac{X}{{{M^4}}}} \right)^{\alpha  - 1}} +
V(\phi), \label{eq:rho}
\\
{p_\phi } &=& X{\left( {\frac{X}{{{M^4}}}} \right)^{\alpha  - 1}} - V(\phi ).\label{eq:p}
\end{eqnarray}
Applying Eqs. (\ref{eq:rho}) and (\ref{eq:p}), the first and second Friedmann equations can be written in the following form
\begin{eqnarray}
\label{eq:Friedmann}
H^{2} &=& \frac{1}{3 M_{\rm p}^2}\left[\left(2\alpha-1\right)X\left(\frac{X}{M^{4}}\right)^{\alpha-1} +\;
V(\phi)\right]~,\label{eq: FR-eqn1}\\\nonumber
\dot{H} &=& -\frac{1}{M_{\rm p}^2}\alpha X\left(\frac{X}{M^4}\right)^{\alpha -1},
\label{eq: FR-eqn2}
\end{eqnarray}
wherein $H\equiv \dot{a}/a $ indicates the Hubble parameter and $M_{\rm p}=1/\sqrt{8\pi G}$
is the reduced Planck mass.

The power-law non-canonical model akin to the canonical Standard Model of inflation gives rise to second order equation of motion \cite{Unnikrishnan:2012}.
Inserting relations  (\ref{eq:rho})-(\ref{eq:p}) in the conservation equation,
${\dot \rho _\phi } + 3H\left( {{\rho _\phi } + {p_\phi }} \right) = 0$, results in the scalar field equation of motion as
\begin{equation}
\label{eq:KG}
\ddot \phi  + \frac{{3H\dot \phi }}{{2\alpha  - 1}}
+ \left( {\frac{{V_{,\phi }}}{{\alpha (2\alpha  - 1)}}}
\right){\left( {\frac{{2{M^4}}}{{{{\dot \phi }^2}}}} \right)^{\alpha
- 1}} = 0,
\end{equation}
where $({,\phi})$ is derivative versus $\phi$. It is notable that, all above equations turns into their canonical counterparts for $(\alpha=1)$.

Here, the first and second  Hubble slow-roll parameters are specified as
\begin{equation}\label{SRP}
  \varepsilon_{1} \equiv -\frac{\dot H}{H^2}, \hspace{.5cm}  \varepsilon_{2} \equiv \frac{\dot{\varepsilon_{1}}}{ H\, \varepsilon_{1}}.
\end{equation}
In the slow-roll approximation regime, the conditions $\{ \varepsilon_{1},\varepsilon_{2}\}\ll 1$ are affirmed and the kinetic energy term can be negligible through the domination of the potential energy. Thereunder Eqs. (\ref{eq:Friedmann})-(\ref{eq:KG}) can be abbreviated \cite{Unnikrishnan:2012} as the ensuing shape

\begin{align}
\label{eq:FR1-SR}
& 3 M_{\rm p}^{2}H^2\simeq V(\phi),\\
  \label{eq:KG-SR}
& \dot \phi  =  -
\theta {\left\{ {\left( {\frac{{{M_{\rm p}}}}{{\sqrt 3 \alpha }}}
\right)\left( {\frac{{\theta V_{,\phi }}}{{\sqrt {V(\phi )} }}}
\right){{\left( {2{M^4}} \right)}^{\alpha  - 1}}}
\right\}^{\frac{1}{{2\alpha  - 1}}}},
\end{align}
wherein $\theta  = 1$ for $ V_{,\phi }>0$ and $\theta  = -1$ for
$ V_{,\phi }<0$. Note that Eq. (\ref{eq:KG-SR}) has been obtained by replacing Eq. (\ref{eq:FR1-SR}) into (\ref{eq:KG}), in which we have neglected the term $\ddot \phi$ in the slow-roll approximation.

Employing the first slow-roll expression $( \varepsilon_{1} \equiv -\dot H/H^2)$ thereto converting the time variable into $e$-folding number $(N)$ via $dN=Hdt$, the combination of Eqs. (\ref{eq:Friedmann})-(\ref{eq:KG}) can lead to the ensuing form of the scalar field equation of motion
\begin{equation}
\label{eq:KG-NC}
\phi_{,NN} + \left[\frac{3}{2\alpha -1} - \varepsilon_{1} \right]\phi_{,N} + \frac{V_{,\phi}}{V}\left[\frac{3\alpha - (2\alpha -1)\varepsilon_{1}}{\alpha(2\alpha-1)} \right]\frac{\phi_{,N}^ {2}}{2\varepsilon_{1}}=0,
\end{equation}
where $(,N)$ and $(,NN)$ indexes denote the first and 2nd derivative with regard to $N$, respectively.

Pursuant to \cite{Garriga:1999}, the slow-roll approximation of the power spectrum of ${\cal R}$ in non-canonical framework at sound horizon spanning $(c_{s}k=aH)$ via comoving wavenumber $k$  can be taken as
\begin{equation}\label{eq:Ps-SR}
{\cal P}_{\cal R}=\frac{H^2}{8 \pi ^{2}M_{\rm p}^{2} c_{s} \varepsilon_{1}}\Big|_{c_{s}k=aH}\,,
\end{equation}
where
\begin{equation}
\label{eq:cs-NC} c_s^2 \equiv \frac{{\partial {p_\phi }/\partial
X}}{{\partial {\rho _\phi }/\partial X}} = \frac{{\partial {\cal
L}(X,\phi )/\partial X}}{{\left( {2X} \right){\partial ^2}{\cal
L}(X,\phi )/\partial {X^2} + \partial {\cal L}(X,\phi )/\partial
X}},
\end{equation}
is the square of sound speed of the scalar perturbations \cite{Garriga:1999,Unnikrishnan:2012}. The imposed observational constraint  by  Planck collaboration  on the amplitude of the curvature power spectrum at pivot scale $(k_{*}=0.05~\rm Mpc^{-1})$  is
 $ {\cal P}_{\cal R}(k_{*})\simeq 2.1 \times 10^{-9}$ \citep{akrami:2018}.

In our non-canonical framework with power-law Lagrangian (\ref{Lagrangian}), one can easily show that the sound speed (\ref{eq:cs-NC}) reads
\begin{equation}
\label{eq:cs}
c_s^2 = \frac{1}{{
2\alpha  - 1 }}.
\end{equation}
Here, to avoid of classical instability we need $c_s^2>0$ which yields $\alpha>1/2$.
The scalar spectral index $n_{s}$ in terms of the slow-roll parameters in non-canonical setup can be derived from the curvature power spectrum as \cite{Garriga:1999}
\begin{align}\label{eq:ns-SR}
n_s-1\equiv \frac{d\ln{\cal P}_{\cal R}}{d\ln k}= -2\varepsilon_1-\varepsilon_2.
\end{align}
The imposed observational constraint  by Planck 2018 on the scalar spectral index  is
$n_s= 0.9668 \pm 0.0037$ (TT,TE,EE+lowE+lensing+BK15+BAO, 68\%  CL)  \cite{akrami:2018}.

As for the tensor perturbation, pursuant to \cite{Garriga:1999}, the slow-roll approximation of the tensor power spectrum at $k=aH$ in the non-canonical setup
can be taken as
\begin{equation}\label{eq:Pt-SR}
{\cal P}_{t}=\frac{2H^2}{\pi ^{2}M_{\rm p}^2}.
\end{equation}
By reason of  dependency of the tensor power spectrum on the gravity term of the action,
equality of the tensor power spectrum in non-canonical model and canonical Standard Model of inflation would be obvious.
Using the scalar  (\ref{eq:Ps-SR}) and tensor (\ref{eq:Pt-SR}) power spectra in non-canonical model, the tensor-to-scalar ratio $r$  could easily be computed  as
\begin{equation}
\label{eq:r}
r\equiv\frac{{\cal P}_t}{{\cal P}_{\cal R}}= 16 c_s \varepsilon_1 .
\end{equation}
The imposed upper limit by Planck 2018 on the tensor-to-scalar ratio   is
$r<0.063$ (TT,TE,EE+lowE+lensing+BK15+BAO, 95\%  CL)  \cite{akrami:2018}. It is notable that, the latest observation by BICE/Keck collaboration has tighten the upper bound on $r$ to $r<0.036$ at $95\%$ CL \cite{BK18:2021}.

%===================================enhanced Curvature Perturbation========================
\section{enhanced curvature perturbations}\label{sec3}
As mentioned previously, so as to create PBHs, a notable increase about seven order of magnitude in the amplitude of ${\cal P}_{\cal R}$ in comparison with $ {\cal P}_{\cal R}(k_{*})$ at CMB scale is necessitated.
This section is devoted to explain how this amplification could take place on small scales, in non-canonical setup with power-law Lagrangian (\ref{Lagrangian}). In Pursuance of this aim, the basic potential of the model is corrected by adding a minute bump like term to make the inflaton slow down on scales smaller than CMB scale. In this way the modified potential is given as
\begin{equation}\label{eq:modified V}
  V(\phi)=V_{b}(\phi)\big[1+\epsilon(\phi)\big],
\end{equation}
wherein $V_{b}(\phi)$ denotes the basic potential in charge of compatibility of the created quantum fluctuations during inflation with the latest CMB observations in $n_{s}$ and $r$. Then, $\epsilon(\phi)$ indicates a minute peaked function of $\phi$ in charge of modifying the basic potential on small scales in order to make the inflaton slow down  and  amplify the scalar power spectrum without disturbing effect on CMB scales. Pursuant to \cite{Rezazadeh:2021,mishra:2020} the bump function is taken as
\begin{equation}\label{eq:bump}
  \epsilon(\phi)=\omega\cosh^{-2}\Big(\frac{\phi-\phi_{c}}{b}\Big),
\end{equation}
which could produce a peak of the height $\omega$ and breadth $b$ in $\phi=\phi_{c}$ position.
The parameters $\{\phi_{c},b\}$ have dimensions of mass, whereas $\omega$ is dimensionless.
 It is worth noting that, for $\phi\neq\phi_{c}$ the effect of bump function lessens and it fades away $(\epsilon(\phi)\ll 1)$.
Tuning the parameters of the bump term $\{\omega,b,\phi_{c}\}$
and $\alpha$ parameter of the Lagrangian (\ref{Lagrangian}) gives rise to amplify ${\cal P}_{\cal R}$ on the necessary scales without significant effect on CMB scale. Ergo, not only the consistency of the model with CMB observation on large scales is guaranteed but also the amplification of ${\cal P}_{\cal R}$ on smaller scales could be occurred.

The quartic potential is considered for the base potential in Eq. (\ref{eq:modified V}) as
\begin{equation}\label{eq:quartic}
V_b(\phi)=\frac{\lambda}{4}\phi^4.
\end{equation}
where $\lambda\simeq 0.13$ is the dimensionless self-coupling constant \cite{Tanabashi:2018}.
It is well understood that, the viable inflationary epoch could not originate from the quartic potential in the Standard Model of inflation. In that case, predictions of this potential for the amplitude of ${\cal P}_{\cal R}$ as well as $n_s$ (at the pivot scale) cannot be consistent with the Planck data \cite{akrami:2018}.
Furthermore, because of the large tensor fluctuation predicted by this potential, its resultant value for  $r$ in canonical standard model of inflation cannot place in the scope of observational data of Planck  2018  \cite{akrami:2018}.  With these aspects in mind, in this work it has been tried to remedy the observational results of quartic potential on CMB scales in power-law non-canonical framework. At the same time,  adding the bump like term to the base potential, creation of detectable PBHs and GWs on smaller scales in this framework has been evaluated.

Thoroughly, this model consists of an assortment of six parameters like $\{\alpha, M,   \lambda, \omega, \phi_c, b\}$. For fixed $\lambda=0.13$, the parameter $M$ can be obtained from the restricted amplitude of the scalar power spectrum  (${\cal P}^{*}_{\cal R}\sim2.1 \times 10^{-9}$) at pivot scale $(k_{*}=0.05~\rm Mpc^{-1})$ \cite{akrami:2018} as  $M=2.65\times10^{-5}M_{\rm p}$. The breadth parameter of the bump for all Cases of the model is set to  $b=7\times10^{-5}M_{\rm p}$. The rest of adjusted parameters
 are listed in Table \ref{tab1}. The numerical results as to inflationary observable values ($n_s$ and $r$) thereto
the  computed  numerical values for  PBHs  formation are summarized in Table \ref{tab2}.
\begin{table}[H]
  \centering
  \caption{The adjusted parameters for Cases A, B, and C.
    Here $\lambda=0.13$, $b=7\times10^{-5}M_{\rm p}$ and $M=2.65\times10^{-5}M_{\rm p}$. }
\begin{tabular}{cccc}
  \hline
 $\#$ &\qquad $\omega$\qquad  & \qquad$\phi_{c}/M_{\rm p}$\qquad&\qquad $\alpha$\qquad \\[0.5ex] \hline\hline
  Case A& \qquad$4.305\times10^{-2}$\qquad  &\qquad$0.01080$ \qquad& \qquad $17.0370470$\qquad \\[0.5ex] \hline
  Case B&\qquad $3.790\times10^{-2}$\qquad &\qquad$0.01173$\qquad & \qquad $17.0735945$\qquad   \\ \hline
  Case C&\qquad$3.459\times10^{-2}$\qquad & \qquad$0.01248$\qquad& \qquad $17.1540680$\qquad \\ \hline
\end{tabular}
 \label{tab1}
\end{table}

\begin{table}[H]
\vspace{-0.6cm}
  \centering
  \caption{The resultant numerical values for the Cases of Table \ref{tab1} as to the scalar spectral index $n_{s}$, the tensor-to-scalar ratio  $r$, the peak values of: the  amplitude of the scalar power spectrum ${\cal P}_{ \cal R}^\text{peak}$, the wavenumber $k_{\text{peak}} $,  the PBHs abundances $f_{\text{PBH}}^{\text{peak}}$ and  masses $M_{\text{PBH}}^{\text{peak}}$ thereto the duration of the inflation
  $\Delta N=N_{\rm end}-N_{*}$.  The values of  $n_{s}$ and $r$ are computed at CMB horizon spanning $e$-fold number $(N_{*}=0)$.}
\begin{tabular}{cccccccc}
  \hline
   $\#$ & \quad $n_{s}$\quad &\quad $r$\quad &\quad$ {\cal P}_{\cal R}^{\text{peak}}$\quad &\quad$k_{\text{peak}}/\text{Mpc}^{-1}$\quad& \quad$f_{\text{PBH}}^{\text{peak}}$\quad& \quad$M_{\text{PBH}}^{\text{peak}}/M_{\odot}$\quad&\quad$\Delta N$\\ \hline\hline
  Case A &\quad0.9623\quad  &\quad0.030\quad& \quad0.042\quad & \quad$3.64\times10^{12}$ \quad&\quad 1\quad &\quad$1.77\times10^{-13}$\quad&\quad 62.75\quad\\ \hline
 Case B &\quad 0.9624\quad  & \quad 0.030\quad &\quad0.051\quad&\quad  $9.68\times10^{8}$ \quad&\quad0.0635\quad &\quad $2.52\times10^{-6}$ \quad&\quad 62.81\quad\\ \hline
 Case C &\quad0.9624\quad  & \quad0.029\quad & \quad0.059\quad & \quad$3.69\times10^{5}$\quad &\quad 0.0012\quad &\quad$17.33$ \quad&\quad 62.91\quad\\ \hline
\end{tabular}
\label{tab2}
\end{table}
\noindent
It is well known that, a duration of 60-70 $e$-folds number of viable inflationary era is needed to amend the shortcomings of Hot Big Bang (HBB) theory \cite{Guth:1981,Liddle:2003,Linde:1983}. For each Case of Table \ref{tab1},  the duration of inflation ($\Delta N=N_{\rm end}-N_{*}$) is defined around 63  $e$-folds (see Table \ref{tab2} for the exact vales), from the CMB horizon spanning moment ($N_{*}=0$) to the end of inflation ($N_{\text{end}}$).
The graph of the first slow-roll parameter $\varepsilon_{1}$ in Fig. \ref{fig:e1,e2,H} indicates that the end of inflation for all Cases of the model occurs at the moment of $\varepsilon_1 =1$.

\begin{figure*}
\centering
\vspace{-0.9cm}
\scalebox{0.6}[0.6]{\includegraphics{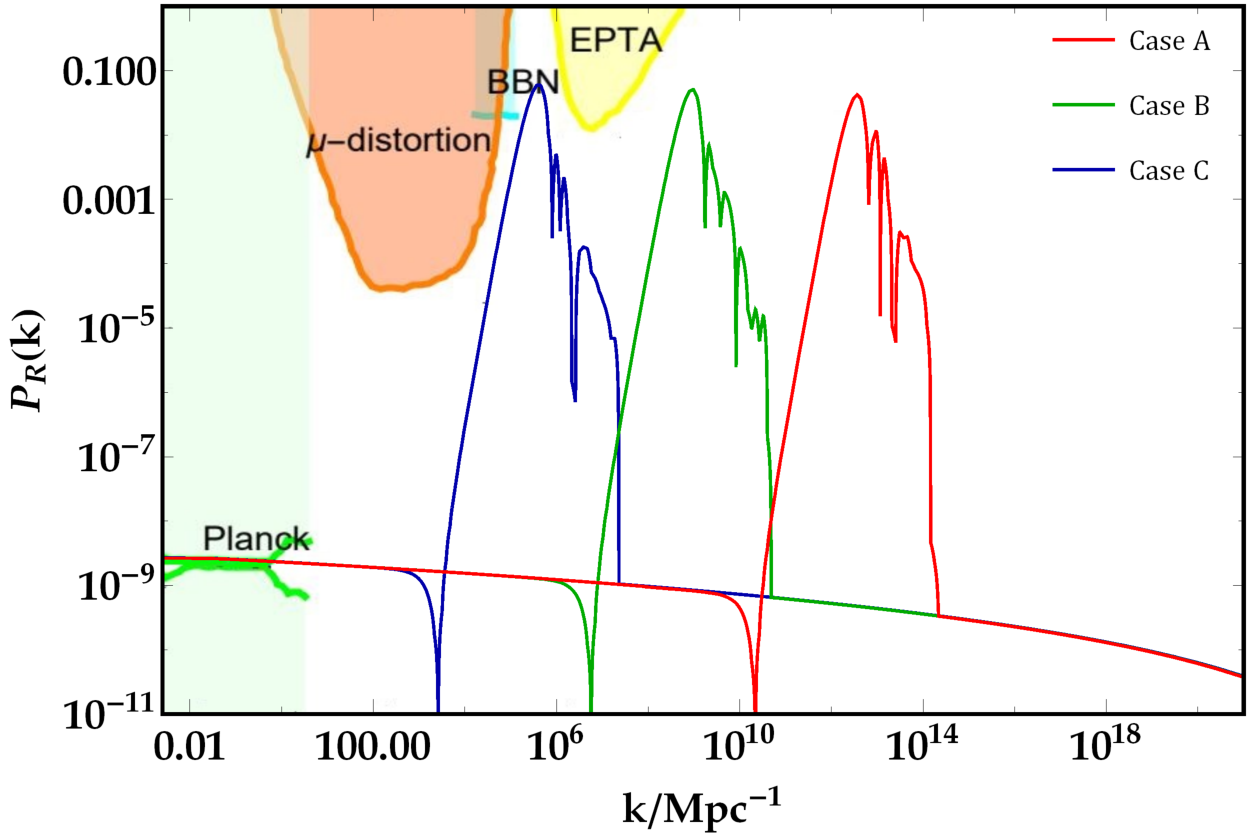}}
\caption{The curvature perturbations power spectra obtained from the exact solution of the MS equation with relation to  comowing  wavenumber $k$ as to the Cases A (red line), B (green line) and C (blue line).  The light-green, yellow, cyan, and orange districts delineate the restrictions of  the CMB observations \cite{akrami:2018},  PTA observations \cite{Inomata:2019-a},   the impact on the ratio between neutron and proton during the Big Bang Nucleosynthesis (BBN) \cite{Inomata:2016,Nakama:2014,Jeong:2014}, and the $\mu$-distortion of CMB \cite{Fixsen:1996,Chluba:2012}, respectively.}
\label{fig:ps}
\end{figure*}

In order to plot the graphs of Fig. \ref{fig:e1,e2,H}, we need to know the dynamical behaviour of the scalar field. Ergo, considering the potential (\ref{eq:modified V}) with the  bump term (\ref{eq:bump}), the second Friedmann equation (\ref{eq: FR-eqn2}) and  the equation of motion (\ref{eq:KG-NC}) are solved simultaneously. The initial conditions for numerical solving of the background equations are taken from the slow-roll approximation (\ref{eq:FR1-SR})-(\ref{eq:KG-SR}) with the basic potential (\ref{eq:quartic}).
In Fig. \ref{fig:e1,e2,H}, the graphs of evolution of field derivative $\phi_{,N}$ thereto  the first and second  slow-roll parameters ($\varepsilon_1$ and $\varepsilon_2$) has been schemed for Cases A (red lines), B (green lines) and C (blue lines). As it can be seen from the $\phi_{,N}$ graph, the velocity of the scalar field in the proximity of the bump position ($\phi=\phi_{c}$) for each case  tends to zero, momentarily. Due to the braking feature of the bump, this reduction in the field velocity takes place. Also at the moment of bump-passing by the scalar field, a high reduction in the first slow roll parameter occurs for each Case of the model (see $\varepsilon_1$ in Fig. \ref{fig:e1,e2,H}). Hence, regarding the Eq. (\ref{eq:Ps-SR}), a significant enhancement in ${\cal P}_{ \cal R}$  is expected owing to the high reduction in the $\varepsilon_1$  at the moment of bump-passing (see Fig. \ref{fig:ps}).

It can be deuced from the graph of $\varepsilon_{2}$ in Fig. \ref{fig:e1,e2,H} that, the slow-roll approximation condition ($\varepsilon_{i}\ll1$) is transgressed by  $\varepsilon_2$ owing to overstepping the limit at the moment of bump-passing for all Cases. Thus, the slow-roll approximation is not valid in the vicinity of the bump position ($\phi=\phi_{c}$),  whereas it is valid at the moment of the CMB horizon passing $N_{*}=0$ for each Case of the model (see Fig. \ref{fig:e1,e2,H}).
Accordingly, computing the scalar spectral index $n_s$ and tensor-to-scalar ratio $r$ using Eqs. (\ref{eq:ns-SR}) and (\ref{eq:r}) in slow-roll approximation is allowed.
The tabulated numerical values in Table \ref{tab2} illustrate that $n_{s}$ and $r$ for all Cases of the model could place in the scope of permitted data by Planck 2018 (TT,TE,EE+lowE+lensing+BK15+BAO, 68\%  CL) \cite{akrami:2018}. It is worth noting that the values of $r$ for all Cases could be consistent with the newest constraint of BICEP/Keck 2018 data ($r<0.036$ at 95$\%$ CL) \cite{BK18:2021}. As a result of considering the quartic potential in the power-law non-canonical model, its observational predictions on CMB scales are revived.

As mentioned formerly, the slow-roll approximation  in the vicinity of bump position is inoperative. So it is not permitted to use Eq. (\ref{eq:Ps-SR}) to evaluate the curvature power spectrum around the bump position. That is why, this equation has been derived under the slow-roll approximation.
Consequently, so as to evaluate the dynamics of the curvature perturbations throughout the inflationary era, numerical solving of the following Mukhanonv-Sasaki (MS) equation is obliged \cite{Garriga:1999}
\begin{equation}\label{eq:MS}
 \upsilon^{\prime\prime}_{k}+\left(c_{s}^2 k^2-\frac{z^{\prime\prime}}{z}\right)\upsilon_k=0,
\end{equation}
\begin{figure*}
\begin{minipage}[b]{1\textwidth}
\vspace{-1cm}
\subfigure{ \includegraphics[width=.6\textwidth]%
{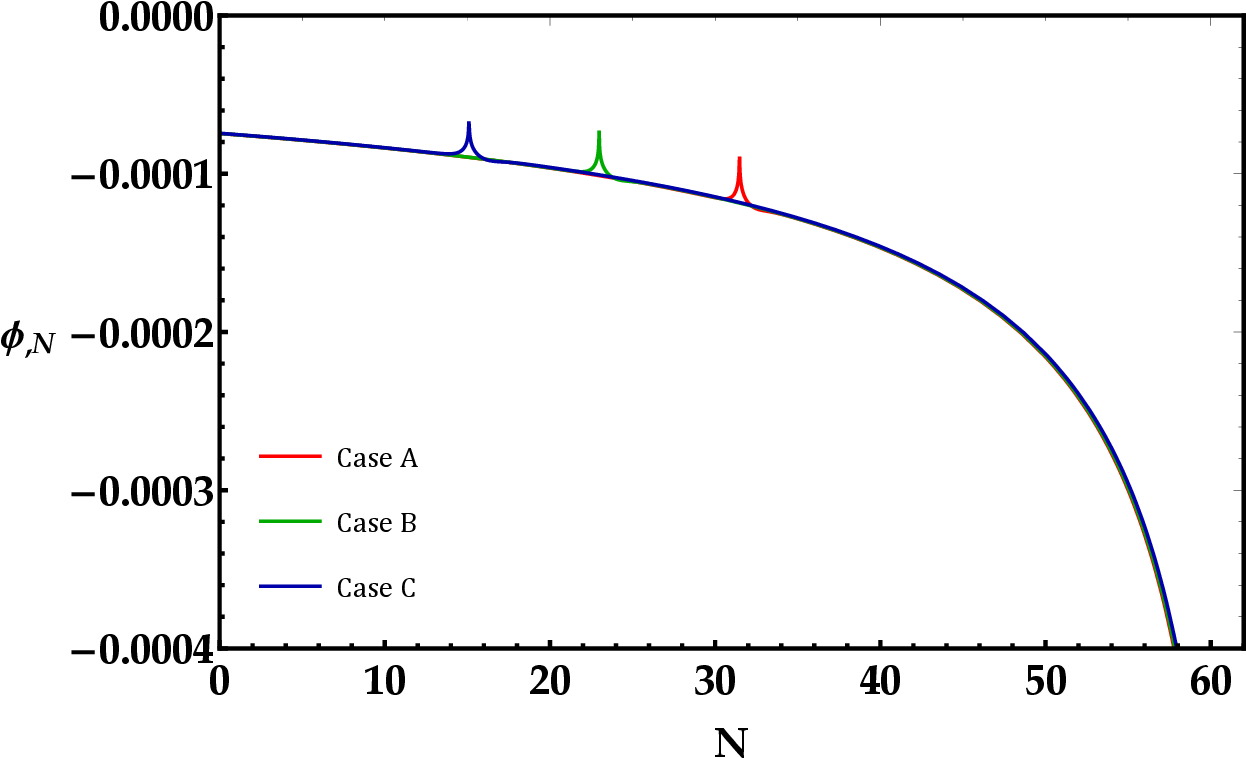}}\hspace{.1cm}
%\subfigure{ \includegraphics[width=.6\textwidth]%
%{H.eps}}
\subfigure{\includegraphics[width=.485\textwidth]%
{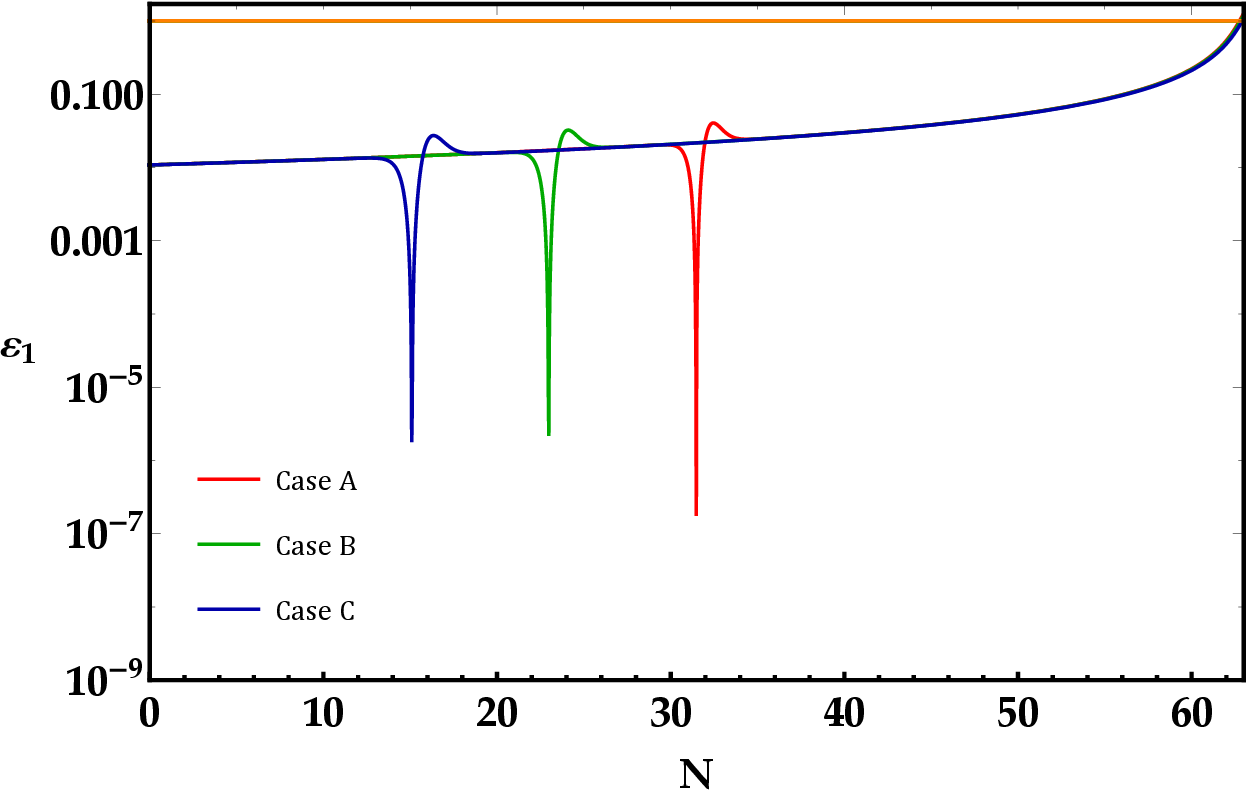}} \hspace{.1cm}
\subfigure{ \includegraphics[width=.485\textwidth]%
{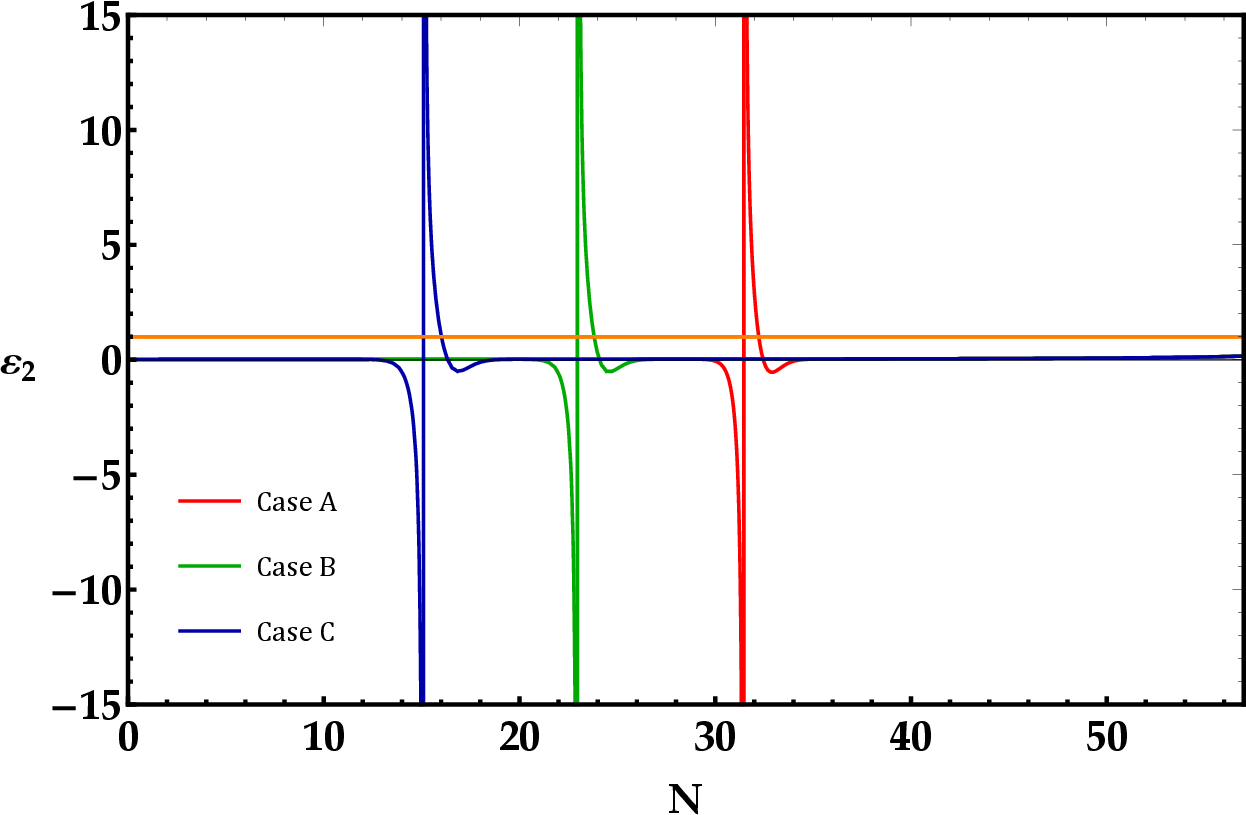}}\hspace{.1cm}
\end{minipage}
\caption{Variations of the field derivative $\phi_{,N}$, first slow-roll parameter $\varepsilon_1$ and second slow-roll parameter $\varepsilon_2$ versus the $e$-folds number $N$ for the Cases A (red lines), B (green lines) and C (blue lines).}
\label{fig:e1,e2,H}
\end{figure*}
wherein prime implies derivative with relation to the conformal time $\eta\equiv\int {a^{-1}dt}$, and
\begin{equation}\label{eq:z}
 \upsilon_k\equiv z {\cal R}_k, \hspace{1cm} z \equiv \frac{a\,\left(\rho_{_{\phi}}+ p_{_{\phi}}\right)^{1/2}}{c_{_s}H}.
\end{equation}
The Bunch-Davies vacuum state at the scales deep inward the horizon can be taken as the nascent condition to solve the MS equation (\ref{eq:MS}) as follows \cite{Garrett:2011}
\begin{equation}\label{eq:Bunch}
\upsilon_k\simeq\frac{e^{-i c_{s}k\eta}}{\sqrt{2c_s k}}, \;\;  (aH\ll c_sk).
\end{equation}
Thence, the power spectrum of the curvature perturbation can be obtained from the exact solution of the MS equation as follows
\begin{equation}\label{eq:PsMS}
{\cal P}_{\cal R}\equiv\frac{k^3}{2\pi^2}\big|{\cal R}_k^2\big|=\frac{k^3}{2\pi^2}\left|{\frac{{\upsilon_k}^2}{z^2}}\right|.
\end{equation}
The accurate values of the amplitudes of the curvature power spectra  ${\cal P}_{\cal R}^{\rm peak}$ and their pertinent comoving wavenumbers $k_{\rm peak}$ at the peak position (corresponding to the bump position ($\phi=\phi_c$) in the potential) for all Cases of the model has been tabulated in Table \ref{tab2}.
Thereafter, in Fig. \ref{fig:ps} the graphs of the computed scalar power spectra ${\cal P}_{\cal R}$ against the comoving wavenumber  $k$ as to all Cases of the model have been mapped. In this figure, beside  the current observational constraints, the ${\cal P}_{\cal R}$ as to the Cases A, B and C have been delineated by red, green and blue solid lines, respectively.
It is deduced from the Fig. \ref{fig:ps} that, in the  CMB scale ($k\sim0.05~ \rm Mpc^{-1}$) proximity placed in slow-roll approximation dominion, the amplitude of curvature power spectra have been confined to the observational constraint
${\cal O}(10^{-9})$ for all Cases. At the same time, in the bump scale proximity (smaller scales than CMB scale), the amplitudes of ${\cal P}_{\cal R}$  experience an enhancement  around $7$ order of magnitude for each Case
and grow up to  ${\cal O}(10^{-2})$. In this way, An adequate circumstance is provided to create PBHs on the bump-passing scales for each Case.

\begin{figure}[H]
\centering
%\vspace{-0.9cm}
\scalebox{0.6}[0.6]{\includegraphics{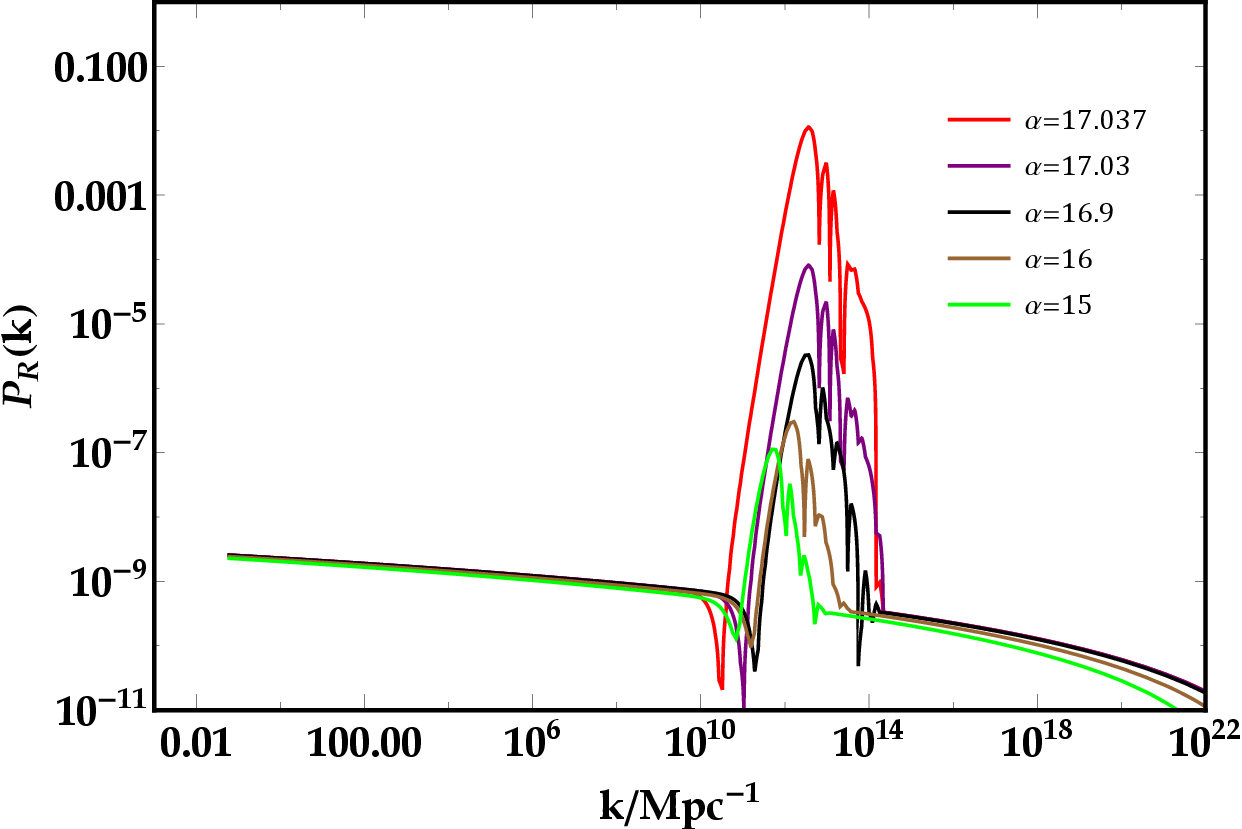}}
%\vspace{-0.6cm}
\caption{The dependence of the curvature perturbations power spectrum on the non-canonical $\alpha$ parameter for Case A of Table \ref{tab1}. The amplitude of ${\cal P}_{\cal R}$ increase with the increase of $\alpha$ in each Case of the model in the bump scales vicinity}
\label{fig:ps_a}
\end{figure}

The notable point about this model is that, the non-canonical $\alpha$ parameter related to the Lagrangian (\ref{Lagrangian}) likewise the bump feature have the enhancing effect on the amplitude of the curvature power spectra on small scales, without disturbing effect on the CMB scale. In other word, the sufficient enhancement in the scalar power spectra to create PBHs could be achievable through fine tuning the bumps parameters as well as the non-canonical $\alpha$ parameter (see Table \ref{tab2} for each Case of the model).
The dependence of the amplitude of the scalar power spectrum for Case A on $\alpha$ parameter is depicted in  Fig. \ref{fig:ps_a}. This figure implies that, the greater amounts of $\alpha$ gives rise to the more enhanced ${\cal P}_{\cal R}$ in the vicinity of the bump scales with no significant effect on ${\cal P}_{\cal R}$ on CMB scale vicinity.
%===================================PBH Mass spectra==================================================
\section{PBHs mass spectrum}\label{sec4}
In this section, creation of PBHs from the amplified perturbed modes, emanated from the quartic potential in the presence of a minute bump in power-law non-canonical setup,  is analyzed. As a consequence of reverting these modes (originated form the inflationary era) to the horizon in RD era, the ultra dense localities have been created. Subsequently,
gravitational collapse of such localities results in create PBHs, as already mentioned.
The mass of these PBHs is specified as a fraction of the horizon as follows
\begin{align}\label{Mpbheq}
M_{\rm PBH}(k)=\gamma\frac{4\pi}{H}\Big|_{c_{s}k=aH} \simeq M_{\odot} \left(\frac{\gamma}{0.2} \right) \left(\frac{10.75}{g_{*}} \right)^{\frac{1}{6}} \left(\frac{k}{1.9\times 10^{6}\rm Mpc^{-1}} \right)^{-2},
\end{align}
where $\gamma=\big(\frac{1}{\sqrt{3}}\big)^{3}$ indicates the collapse efficiency \cite{carr:1975} and  $g_{*}=106.75$ is the effective number of relativistic degrees of freedom.
Surmising the Gaussian distribution for the curvature perturbations, the creation rate of PBHs with mass $M(k)$ using the  Press-Schechter formalism is calculated as \cite{Tada:2019,young:2014}
\begin{equation}\label{betta}
  \beta(M)=\int_{\delta_{c}}\frac{{\rm d}\delta}{\sqrt{2\pi\sigma^{2}(M)}}e^{-\frac{\delta^{2}}{2\sigma^{2}(M)}}=\frac{1}{2}~ {\rm erfc}\left(\frac{\delta_{c}}{\sqrt{2\sigma^{2}(M)}}\right),
\end{equation}
in which "erfc" is the error function complementary and $\delta_{c}$ indicates the density perturbations threshold specified as $\delta_{c}=0.4$ \cite{Musco:2013,Harada:2013}. Here,  $\sigma^{2}(M)$ is the coarse-grained density contrast smoothed on the scale $k$ given by
\begin{equation}\label{sigma}
\sigma_{k}^{2}=\left(\frac{4}{9} \right)^{2} \int \frac{{\rm d}q}{q} W^{2}(q/k)(q/k)^{4} {\cal P}_{\cal R}(q),
\end{equation}
wherein ${\cal P}_{\cal R}$ is the curvature power spectrum, and  $W$ is the Gaussian window $W(x)=\exp{\left(-x^{2}/2 \right)}$.
In the end, the  abundance of PBHs is specified as the ensuing form
\begin{equation}\label{fPBH}
f_{\rm{PBH}}(M)\simeq \frac{\Omega_{\rm {PBH}}}{\Omega_{\rm{DM}}}= \frac{\beta(M)}{1.84\times10^{-8}}\left(\frac{\gamma}{0.2}\right)^{3/2}\left(\frac{g_*}{10.75}\right)^{-1/4}
\left(\frac{0.12}{\Omega_{\rm{DM}}h^2}\right)
\left(\frac{M}{M_{\odot}}\right)^{-1/2},
\end{equation}
where, according to the Planck 2018 data \cite{akrami:2018}, the current DM density parameter is specified as $\Omega_{\rm {DM}}h^2\simeq0.12$.

\begin{figure}[H]
\centering
\includegraphics[scale=0.6]{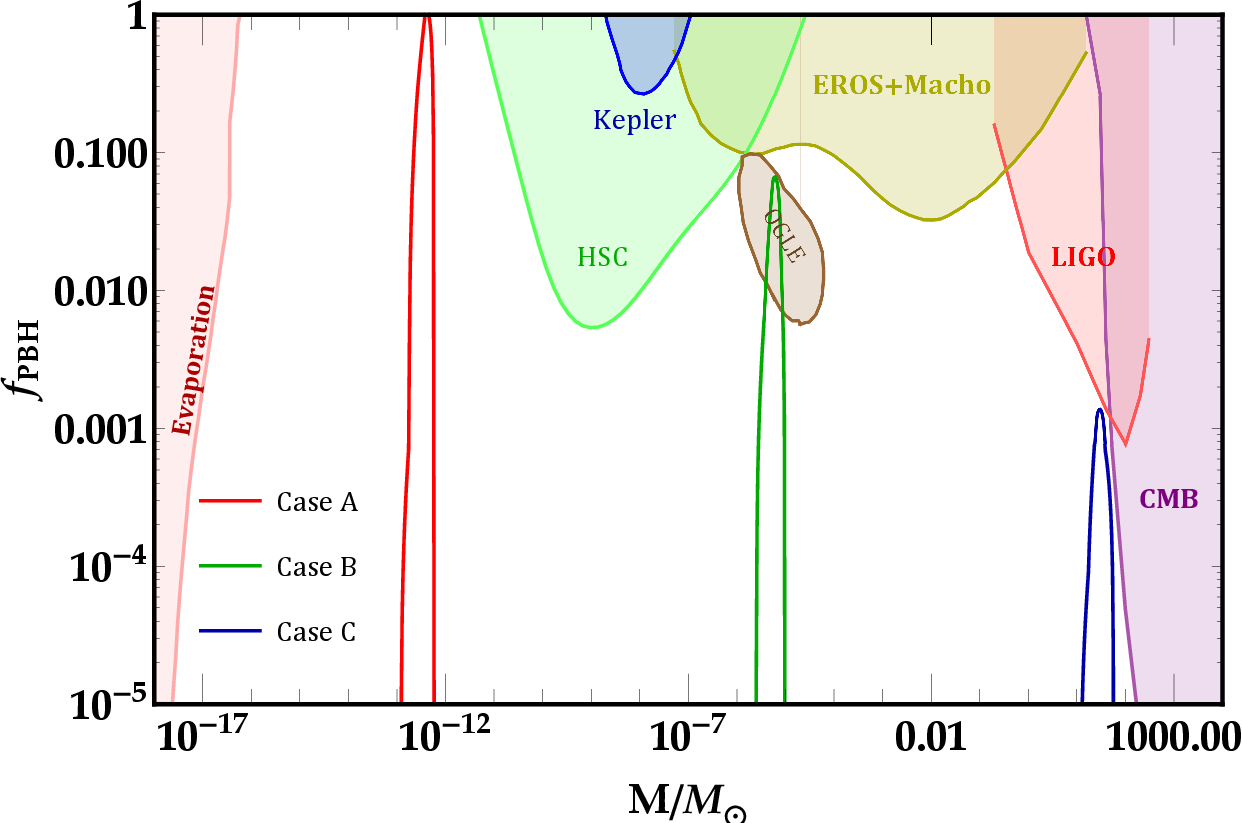}
\caption{The PBHs mass spectra as to the Cases A (red line), B (green line) and C (blue line).  The colorful sectors portray the current confinement on the PBHs abundance. Most of these sectors are prohibited by constraints of CMB \cite{CMB} (purple sector), LIGO-VIRGO event \cite{Abbott:2019,Chen:2020,Boehm:2021,Kavanagh:2018} (red sector), microlensing events via MACHO \cite{MACHO}, EROS \cite{EORS}, Kepler \cite{Kepler}), Icarus \cite{Icarus}, OGLE \cite{OGLE-1,OGLE-2},  and Subaru-HSC \cite{subaro} (green sector), PBHs evaporation \cite{EGG, Laha:2019,Clark,Shikhar:2022,Dasgupta} (pink sector). The only  permitted sector is the ultrashort-timescale microlensing events in the OGLE data \cite{OGLE-1,OGLE-2} (brown sector).}
\label{fig-fpbh}
\end{figure}

Consequently, replacing the numerical values of the ${\cal P}_{\cal R}$ from solving the  MS Eq. (\ref{eq:MS}) in Eq. (\ref{sigma}) and employing  Eqs. (\ref{Mpbheq})-(\ref{fPBH}), the PBHs abundance for each Case of Table \ref{tab1} can be calculated. Thence, the numerical resultant values have been tabulated in Table \ref{tab2} and the graphs of PBHs mass spectra have been plotted in Fig. \ref{fig-fpbh} for all Cases of the model.
The foretold PBHs for the parameter Case A with a mass spectrum located in asteroid-mass scope through $M_{\rm PBH}^{\rm peak}=1.77\times10^{-13}$ could be an acceptable candidate for the whole DM content ($f_{\rm PBH}^{\rm peak}\simeq1$). The acquired PBHs mass spectrum for the parameter Case B has placed in the earth-mass scope through  $M_{\rm PBH}^{\rm peak}=2.52\times10^{-6}$ and abundance  $f_{\rm PBH}^{\rm peak}=0.0635$. The peak of the mass spectrum of this case has situated  in the permitted sector of the OGLE data \cite{OGLE-1,OGLE-2}, therefore it could be appropriate to explain the ultrashort-timescale microlensing events. As for the parameter Case C, the resultant PBHs in the stellar-mass scope with  $M_{\rm PBH}^{\rm peak}=17.33M_{\odot}$ and $f_{\rm PBH}^{\rm peak}\simeq0.0012$ have located in the observable band of LIGO-VIRGO events, hence their coeval GWs can be detected by these detectors.

It is worth noting that, in order to compute the PBHs abundance $f_{\rm PBH}$ in this section, we have considered the Gaussian distribution for the scalar perturbations. Nevertheless, it is proven by Atal et al. \cite{Atal:2019,Atal:2020} that, when the inflaton potential is considered in the presence of a bump,  the non-attractor evolution around the bump will make the final curvature perturbation non-Gaussian.
Moreover, it is well known that, the non-canonical framework with power-law Lagrangian
leads to the equilateral type of the non-Gaussianity parameter which  depends on $\alpha$ parameter as $f_{\rm NL}^{\rm equil}=-\frac{275}{486}(\alpha-1)$ \cite{Rezazadeh:2015}. So, $f_{\rm NL}^{\rm equil}\simeq-9$ for all Cases of Table \ref{tab1}, which could be consistent with the constraints of Planck 2018 on primordial non-Gaussianity ($f_{\rm NL}^{\rm equil}=-26\pm47$ 68\%  CL) \cite{akrami-non:2018}. In \cite{Atal:2019,Atal:2020}, it is inferred that, the  PBHs abundance could be amplified through taking the non-Gaussianity into account. It means that, one could catch the favorable abundances for PBHs with the smaller amplitude of the scalar power spectrum than ${\cal O}(10^{-2})$ \cite{Atal:2019,Atal:2020}.
However, we have not taken the effect of non-Gaussianity  (brought by the bump in the potential
and non-canonical kinetic term) into account on the PBHs abundance in this work and we postpone the analysis of this issue to future works.
%===================================secondary Gravitational waves=========================================
\section{ secondary gravitational waves}\label{sec5}
The secondary GWs may be emanated from the re-entering amplified scalar perturbation modes coeval with PBHs creation in RD stage. These GWs can be traceable through the various earth-based or space-based detectors, if their frequencies be situated in the sensibility bands of them. The traceability of the GWs signals can afford the researchers to check the rectitude of the multifarious inflationary models. That is why, probing the secondary GWs in PBHs concepts has been an interesting topic theses days. This section is  allotted to compute the current energy density spectra of the secondary GWs in terms of frequency as to the created PBHs (Fig. \ref{fig-fpbh}) from the quartic potential with a tiny bump in the power-law non-canonical setup.
The  energy density of the secondary GWs in RD stage is formulated as \cite{Kohri:2018}
\begin{eqnarray}\label{OGW}
&\Omega_{\rm{GW}}(\eta_c,k) = \frac{1}{12} {\displaystyle \int^\infty_0 dv \int^{|1+v|}_{|1-v|}du } \left( \frac{4v^2-(1+v^2-u^2)^2}{4uv}\right)^2\mathcal{P}_{\cal R}(ku)\mathcal{P}_{\cal R}(kv)\left( \frac{3}{4u^3v^3}\right)^2 (u^2+v^2-3)^2\nonumber\\
&\times \left\{\left[-4uv+(u^2+v^2-3) \ln\left| \frac{3-(u+v)^2}{3-(u-v)^2}\right| \right]^2  + \pi^2(u^2+v^2-3)^2\Theta(v+u-\sqrt{3})\right\}\;,
\end{eqnarray}
wherein  $\Theta$ and $\eta_{c}$ signify the Heaviside theta function, and the increment ending time of the $\Omega_{\rm{GW}}$.
The association between the present GWs energy spectrum and its counterpart at $\eta_{c}$
is expressed by  \cite{Inomata:2019-a}
\begin{eqnarray}\label{OGW0}
\Omega_{\rm GW_0}h^2 = 0.83\left( \frac{g_{*}}{10.75} \right)^{-1/3}\Omega_{\rm r_0}h^2\Omega_{\rm{GW}}(\eta_c,k)\;,
\end{eqnarray}
where $\Omega_{\rm r_0}h^2\simeq 4.2\times 10^{-5}$ denotes the present radiation density parameter and $g_{*}\simeq106.75$ defines the effective degrees of freedom in the energy density at $\eta_c$. The frequency is pertained to wavenumber by way of
\begin{eqnarray}\label{k_to_f}
f=1.546 \times 10^{-15}\left( \frac{k}{{\rm Mpc}^{-1}}\right){\rm Hz}.
\end{eqnarray}

\begin{figure}[H]
\centering
\includegraphics[scale=0.5]{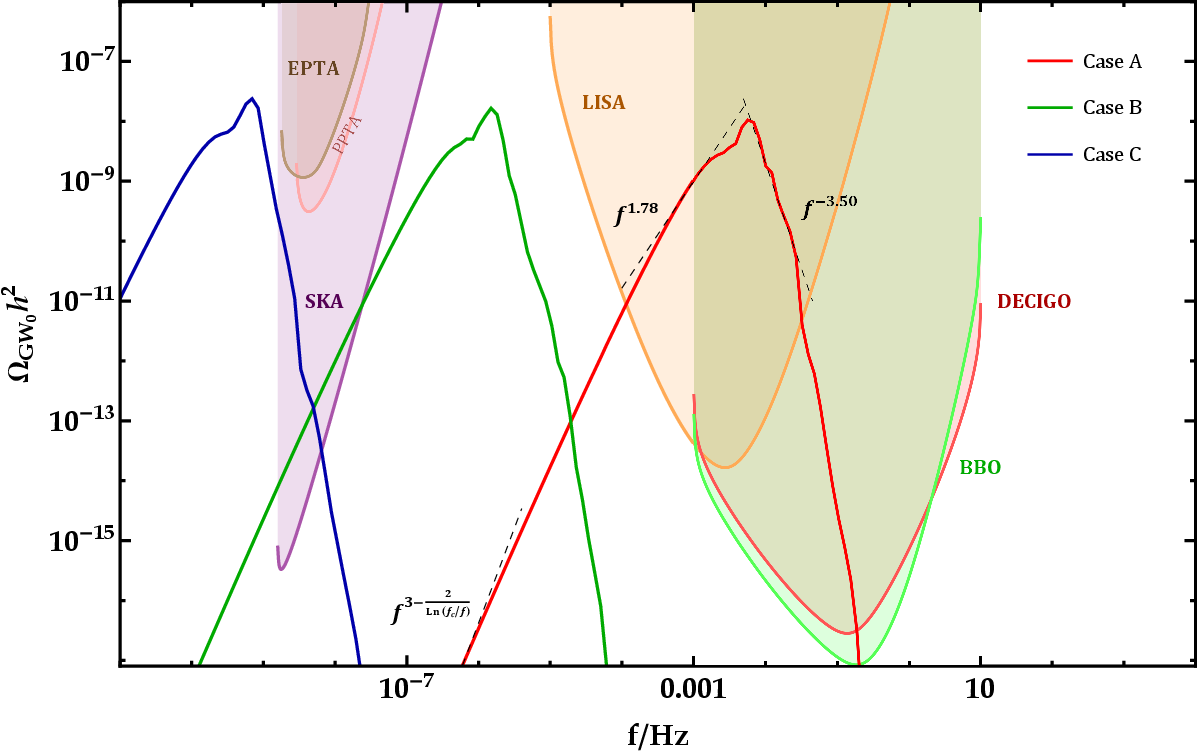}
\vspace{-0.5em}
\caption{ The resultant  present energy density spectra of secondary gravitational waves  $\Omega_{\rm GW_0}$ in relation to frequency as to the parameter Cases A (red line), B (green line) and C (blue line) of Table \ref{tab1}, thereto the observational domains of the GWs detectors like EPTA (brown domain), SKA (purple domain), LISA (orange domain), DECIGO (red domain) and BBO (green domain).  The  power-law aspect of $\Omega_{\rm GW_0}$ is portrayed by dashed black lines in three frequency zones for Case A.}
\label{fig-omega}
\end{figure}

In this stage, utilizing the numerical values of ${\cal P}_{\cal R}$ derived from the precise  solution of the MS equation (\ref{eq:MS}), thereto Eqs. (\ref{OGW})-(\ref{k_to_f}) the present energy  spectra of  secondary GWs as to  PBHs for each Case of Table \ref{tab1} are calculated.
The resultant graphs have been verified in the light of sensibility bands of GWs detectors such as  SKA (purple domain) \cite{ska}, EPTA (brown domain) \cite{EPTA-a,EPTA-b,EPTA-c,EPTA-d}, LISA (orange domain) \cite{lisa,lisa-a},  BBO (green domain) \cite{Yagi:2011,BBO:2003} and DECIGO (red domain) \cite{Yagi:2011,Seto:2001}  (see Fig. \ref{fig-omega}).
It can be inferred from this figure that,  the resultant spectra of $\Omega_{\rm GW0}$ as to the parameter Cases A (red line), B (green line), C (blue line) have  peaks of nearly identical height about $10^{-8}$ placed in diverse frequencies (see Table \ref{table:GWs}).
The frequency of the peak of $\Omega_{\rm GW_0}$ spectrum  as to Case A is around ${\cal O}(10^{-3})$, and it can be tracked down by LISA. Moreover, the frequencies of the peaks of $\Omega_{\rm GW_0}$ spectra as to Cases B and C are around ${\cal O}(10^{-6})$ and ${\cal O}(10^{-10})$, respectively and they can be traced via SKA detector. As a result, the verity of this model could be checked by dint of the forthcoming data of these detectors.
In the end, the inclinations of the $\Omega_{\rm GW_0}$ spectra at various frequency bands have been appraised . It has been demonstrated that, the current density spectra of secondary GWs have a power-law inclinations in terms of frequency as  $\Omega_{\rm GW_0} (f) \sim f^{n} $ \cite{fu:2020,Xu,Kuroyanagi}.
Table \ref{table:GWs} embodies the computed frequencies of peaks ( $f_{c}$) and appraised power index $n$ in three frequency bands as $f\ll f_{c}$,  $f<f_{c}$ and  $f>f_{c}$ for all Cases of our model. As another consequence, the resultant values for $n$ in the infrared band $f\ll f_{c}$ can be consistent with the logarithmic relation $n=3-2/\ln(f_c/f)$  procured in \cite{Yuan:2020,shipi:2020,Yuan:2023}.
\begin{table}[ht!]
  \centering
  \caption{The frequencies and heights of the peaks of $\Omega_{\rm GW_0}$, thereto the power index $n$ in frequency bands $f\ll f_{c}$,  $f<f_{c}$ and  $f>f_{c}$ as to Cases A, B and C.}
\scalebox{1}[1] {
\begin{tabular}{cccccc}
\hline
\#  & $\qquad\qquad$ $f_{c}$ $\qquad\qquad$ & $\quad$ $\Omega_{\rm GW0}\left(f_{c}\right)$ $\quad$ & $\quad$ $n_{f\ll f_{c}}$ $\quad$ & $\quad$ $n_{f<f_{c}}$ $\quad$ & $\quad$ $n_{f>f_{c}}$\tabularnewline
\hline
\hline
Case A & $5.642\times10^{-3}$ & $1.049\times10^{-8}$ & $3.011$ & $1.784$ & $-3.503$ \tabularnewline
\hline
Case B & $1.497\times10^{-6}$ & $1.654\times10^{-8}$ & $3.018$ & $1.244$ & $-1.780$\tabularnewline
\hline
Case C & $6.945\times10^{-10}$ & $2.378\times10^{-8}$ & $2.883$ & $1.133$ & $-1.999$\tabularnewline
\hline
\end{tabular}
    }
  \label{table:GWs}
\end{table}
%========================================Conclusions=====================================================
\section{Conclusions}\label{sec6}
In this work,  PBHs creation form the quartic potential in the presence of a minute bump has been investigated. The non-canonical scalar field inflationary framework with a power-law lagrangian (\ref{Lagrangian}) has been chosen so as to not only have a viable inflation emanated from the quartic potential, but also be able to create detectable PBHs.

Inconsistency between the prognostications of the quartic potential  (for $n_s$ and $r$) and observational data on CMB scale \cite{akrami:2018} in the Standard Model of inflation is well known. That is why, this potential is considered in the power-law non-canonical setup to emanate viable inflationary era.
With regard to the Table \ref{tab2}, the resultant values of scalar spectral index $n_s$ and tensor-to-scalar ratio $r$ for all Cases of this model lie inside the permitted data of Planck 2018 (TT,TE,EE+lowE+lensing+BK15+BAO, 68\%  CL) \cite{akrami:2018}.
Also the values of $r$ for all Cases of the model satisfy the newest constraint $r<0.036$ of BICEP/Keck 2018 data at 95$\%$ CL \cite{BK18:2021}.
As a consequence of choosing the power-law non-canonical setup, the observational predictions of the quartic potential could be rectified.

The presence of the minute bump (\ref{eq:bump}) in the quartic potential takes effect as a brake and makes the inflaton slow down on small scales momentarily, without a significant effect on large scales (CMB scale). Ergo, fine tuning of the bump parameters and the non-canonical $\alpha$ parameter of the Lagrangian (\ref{Lagrangian}) (see Table \ref{tab1}) give rise to slow the inflaton down in a moment. In this way, enough time to enhance the curvature perturbations in the bump position on small scales is provided, without disturbing the CMB scale.

After solving the background Eqs. (\ref{eq: FR-eqn2})-(\ref{eq:KG-NC}) numerically, evolution of the field velocity $\phi_{,N}$, the first and second slow-roll parameters ($\varepsilon_1$ and $\varepsilon_2$) against the $e$-fold number $N$ in Fig. \ref{fig:e1,e2,H} have been delineated. Regarding this figure, at the moment of bump-passing  $\varepsilon_1$ experiences a high reduction and guarantees the enhancement in the scalar perturbation modes (\ref{eq:Ps-SR}). At the bump-passing moment the $\varepsilon_1$ comports the slow-roll condition ($\varepsilon_i\ll1$) but the $\varepsilon_2$ violates it.
Hence, the power spectra of the curvature perturbations as to all Cases of Table \ref{tab1} have been obtained through the numerical solutions of the  Mukhanov-Sasaki equation (see Table \ref{tab2} and Fig.\ref{fig:ps}). The depicted scalar power spectra in Fig. \ref{fig:ps} take nearly fixed values compatible with the Planck 2018 data on CMB scales, although on smaller scales they exhibit a peak of sufficient height to produce detectable PBHs.

Using the attained scalar power spectra from the numerical solution of Mukhanov-Sasaki equation and  Press-Schechter formalism, the PBHs mass spectra for each  Case of Table \ref{tab1} have been analyzed. The foretold PBHs for the parameter Case A with mass $M_{\rm PBH}^{\rm peak}=1.77\times10^{-13}$ and $f_{\rm PBH}^{\rm peak}\simeq1$ could be considered as an acceptable candidate for the whole DM content. The acquired PBHs for the parameter Case B with $M_{\rm PBH}^{\rm peak}=2.52\times10^{-6}$ and abundance  $f_{\rm PBH}^{\rm peak}=0.0635$  has situated  in the permitted sector of the OGLE data \cite{OGLE-1,OGLE-2}, therefore it could be appropriate to explain the ultrashort-timescale microlensing events. The resultant PBHs for the parameter Case C with  $M_{\rm PBH}^{\rm peak}=17.33M_{\odot}$ and $f_{\rm PBH}^{\rm peak}\simeq0.0012$ have located in the observable band of LIGO-VIRGO events, hence their coeval GWs can be detected by these detectors (see Table \ref{tab2} and Fig. \ref{fig-fpbh}).

In the following, the propagated secondary GWs originated from PBHs creation have been evaluated. Thence, the spectra of $\Omega_{\rm GW_0}$ for the each Case of Table \ref{tab1} have been computed.
The resultant graphs of $\Omega_{\rm GW0}$ as to the parameter Cases A, B, C  have peaks of nearly identical height about $10^{-8}$ placed in diverse frequencies (see Table \ref{table:GWs}) and they have been verified in the light of sensibility bands of GWs detectors. The $\Omega_{\rm GW_0}$ spectrum  as to Case A can be tracked down by LISA, whereas
the $\Omega_{\rm GW_0}$ spectra as to Cases B and C can be traced via SKA detector (see Fig.\ref{fig-omega}). As a result, the verity of this model could be checked by dint of the forthcoming data of these detectors.

Finally, the power-law inclinations of the $\Omega_{\rm GW_0}$ spectra ($\Omega_{\rm GW_0} (f) \sim f^{n} $) \cite{fu:2020,Xu,Kuroyanagi} at various frequency bands have been appraised for all Cases of Table \ref{tab1} (see table \ref{table:GWs} for appraised power index $n$ for each Case).
As the last consequence, the resultant values for $n$ in the infrared band $f\ll f_{c}$ for all Cases can be consistent with the logarithmic relation $n=3-2/\ln(f_c/f)$  procured in \cite{Yuan:2020,shipi:2020,Yuan:2023}.
%========================================Acknowledgements=====================================================
\section{Acknowledgements}\label{sec7}
This work has been  supported financially by Vice President for Research and Technology, University of Kurdistan.

%===========================================Refrence======================================================

\end{document}